\documentstyle[psfig,11pt,aaspp4]{article}

\received{}
\accepted{}
\journalid{}{}
\articleid{}{}

\slugcomment{Submitted to {\it The Astrophysical Journal Letters}}

\lefthead{Halpern et al.}
\righthead{Optical Afterglow of GRB~980519}

\begin{document}

\title{The Rapidly Fading Optical Afterglow of GRB~980519}

\author{J. P. Halpern\altaffilmark{}, J. Kemp, and T. Piran\altaffilmark{1,2}}
\affil{Astronomy Department, Columbia University, 550 West 120th Street,
    New York, NY 10027}

\and

\author{M. A. Bershady\altaffilmark{}}
\affil{Astronomy Department, University of Wisconsin, 475 N. Charter Street,
    \\Madison, WI 53706}

\altaffiltext{1}{Racah Institute of Physics, The Hebrew University, Jerusalem 91904, Israel.} 
\altaffiltext{2}{Department of Physics, New York University, New York, NY 
10003.} 

\begin{abstract}
GRB~980519 had the most rapidly fading of the well-documented GRB afterglows,
consistent with $t^{-2.05 \pm 0.04}$ in $BVRI$ as well as in X-rays during
the two days in which observations were made.  We 
report $VRI$ observations from the MDM 1.3m and WIYN 3.5m telescopes,
and we synthesize an optical spectrum from all of the available photometry.
The optical spectrum alone is well fitted by a power law of the form
$\nu^{-1.20 \pm 0.25}$, with some of the uncertainty due to the significant
Galactic reddening in this direction.  The optical and X-ray spectra together
are adequately fitted by a single power law $\nu^{-1.05 \pm 0.10}$.
This combination of steep temporal decay and flat broad-band spectrum
places a severe strain on
the simplest afterglow models involving spherical blast waves in a 
homogeneous medium.  Instead, the rapid observed
temporal decay is more consistent with models of expansion into
a medium of density $n(r) \propto r^{-2}$,
or with predictions of the evolution of a jet
after it slows down and spreads laterally.
The jet model would relax the
energy requirements on some of the more extreme GRBs, of which
GRB~980519 is likely to be an example because of its large 
$\gamma$-ray fluence and faint host galaxy.
\end{abstract}

\section{Introduction}

The localization of gamma-ray bursts (GRBs) by
the Wide Field Camera (WFC) on the BeppoSAX satellite (Boella et al. 1997)
has enabled rapid and detailed follow-up 
studies to be made at other wavelengths, including x-ray
(Costa et al. 1997), optical (van Paradijs et al. 1997),
and radio (Frail et al. 1997).  Perhaps the most important result
of this breakthrough has been the measurement of cosmological
redshifts for five bursts (Metzger et al. 1997; Kulkarni et al. 1998, 1999;
Djorgovski et al. 1998a, 1999), and the detection of faint host galaxies for
these five plus four others, leading to the conclusion that the majority of,
if not all GRBs, are the most energetic events in the Universe.

GRB~980519 was one of the brightest in the
BeppoSAX WFC (Muller et al. 1998; in 't Zand et al. 1999),
second only to GRB 990123 (Feroci et al. 1998).
The onset of 2--27 keV emission from GRB 980159 in the WFC
preceded the BATSE trigger on May 19.514035 UT
by about 70~s (in 't Zand et al. 1999), a phenomenon which
characterizes only a few percent of bursts.  The BATSE measured
fluence above 25~keV was $(2.54 \pm 0.41) \times 10^{-5}$ ergs~cm$^{-2}$,
among the top 12\% of BATSE bursts (Connaughton 1998).
An X-ray observation with the BeppoSAX Narrow Field Instruments (NFI)
began 9.7 hr after the burst (Nicastro et al. 1998), and detected an
afterglow with a
2--10~keV flux of $(3.8 \pm 0.6) \times 10^{-13}$ ergs~cm$^{-2}$~s$^{-1}$
in the first 4 hours, which faded by about a factor of 4
in the 1.2 days following.

The optical afterglow of GRB~980519 was discovered by
Jaunsen et al. (1998) 8.8~hr after the burst
using the 2.5m Nordic Optical
Telescope (NOT).  Initially estimated at magnitude $I = 19.5$,
this was the first optical afterglow to appear brighter
than the limiting magnitude of the Palomar Sky Survey,
thus enabling immediate recognition. Djorgovski et al. (1998b)
confirmed the optical transient (OT), and noted that its decay was
consistent with $t^{-1.98}$.
Kemp and Halpern (1998) obtained additional photometry and calibrations
which showed that the initial NOT detection was actually brighter than
first suggested ($I = 18.4$, see below).  VLA observations 
within the first three days (Frail, Taylor, \& Kulkarni 1998)
detected a variable 8.3 GHz source at the position
of the OT.  Sub-millimeter observations
at the JCMT eight days after the burst yielded only upper limits
(Smith et al. 1999).  Two months after the burst, deep optical observations
detected a faint coincident object of magnitude
$R = 26.05 \pm 0.22$ (Sokolov et al. 1998) or $R = 26.1 \pm 0.3$
(Bloom et al. 1998a), which is presumed to be the GRB host galaxy.
Its redshift has not been determined.

\section{Optical Observations}

We obtained $I$ and $R$ photometry of GRB~980519
on the MDM Observatory 1.3m telescope,
and a $V$ image on the WIYN 3.5m telescope, both on Kitt Peak.
Figure~1 shows the $R$-band image and the BeppoSAX
NFI error circle.  We measure a position for the OT
of (J2000) $23^{\rm h}22^{\rm m}21.\!^{\rm s}49,
+77^{\circ}15^{\prime}43.\!^{\prime\prime}3$, with an error radius
of $0.\!^{\prime\prime}35$.  This is consistent with the radio position
reported by Frail et al. (1998).  Exposure times were
$6\times 600$~s in $I$ and $6\times 300$~s in $R$.  The sky was
reasonably clear during the MDM observations, where we also obtained
calibrations using the standard star field of PG~0918+029 (Landolt 1992).
Conditions started
to deteriorate soon thereafter, when a single uncalibrated 600~s $V$
image was obtained at WIYN.  The results of these observations are 
listed in Table~1.  Also listed are all available photometry from
other observatories.  Some of these observations were obtained with
independent calibrations, and some were reported with respect to
comparison stars calibrated by others.  We have done our best to
convert all of the reported magnitudes to a common system in Table~1.
During this campaign three different calibrations were employed,
those of MDM (Kemp \& Halpern 1998), Palomar (Bloom et al. 1998b, Gal et al.
1998), and the USNO (Henden et al. 1998).  Since the USNO calibration
set is the most complete, and was obtained on photometric nights with Landolt
standards, we have chosen to adopt
this system.  We find that in $I$ and $R$, the MDM magnitudes of
comparison stars are
fainter than the USNO by 0.10, while the Palomar magnitudes are fainter
by 0.07.  In $V$ the Palomar magnitudes are brighter than the USNO
by 0.2, while in $B$ Palomar is brighter by 0.14.  Accordingly, these
corrections have been applied to the published magnitudes.

In Figure~2 we graph the light curves in four bands using the data
from Table~1.  For clarity, upper limits are not plotted as they
do not contribute any additional
constraints.  Also not plotted are the unfiltered magnitudes of
Maury et al. (1998), although these are roughly consistent with the
$R$-band light curve.  All of the bands are consistent with the
same power-law decay, $t^{-2.05\pm 0.04}$.  The solid lines represent the
best fit to a single decay constant,
with only the normalization adjusted for each band.
We note that the power-law decay index of the X-ray afterglow,
$\delta_{\rm x} = -2.07 \pm 0.11$ as reported by Owens et al. (1998), is
consistent with the optical, and that these are cotemporaneous
observations.

\section{Continuum Shape and Reddening}

It is possible to synthesize a $BVRI$ spectrum from these data by interpolating
the magnitudes to a particular time.  We chose a time of May 20.34 UT,
19.8~hr after the burst,
because it coincides with the largest number of
measurements, and because the resulting spectrum can be compared
with the X-ray flux from the simultaneous BeppoSAX follow-up.
The interpolated $BVRI$ magnitudes were converted to fluxes using the
effective wavelengths and normalizations of Fukugita, Shimasaku, \&
Ichikawa (1995), and graphed as filled circles in Figure~3.  Galactic
reddening is a significant factor in this field because of its
intermediate Galactic latitude, $(\ell,b) = (117.\!^{\circ}963,
+15.\!^{\circ}285)$.
The selective extinction $E(B-V)$ can be estimated in at least two
ways.  First is the value of Schlegel, Finkbeiner, \& Davis (1998)
from the {\it IRAS\/} $100\mu$m maps, $E(B-V) = 0.267$~mag.  This is somewhat
different from a second estimate, $E(B-V) = 0.348$~mag, which can be
derived from the Galactic 21 cm column density in this
direction, $N_{\rm HI} = 1.74 \times 10^{21}$~cm$^{-2}$
(Stark et al. 1992), and the standard
conversion $N_{\rm HI}/E(B-V) = 5.0 \times 10^{21}$~cm$^{-2}$~mag$^{-1}$
(Savage \& Mathis 1979).  Figure~3 shows the results of applying
each of these corrections, using the relative extinctions from
Schlegel et al. (1998). In either case, the spectrum is a good
fit to a power law of the form $F_{\nu} \propto \nu^{\alpha}$.
The smaller extinction requires $\alpha = -1.25 \pm 0.20$, while the
larger extinction corresponds to $\alpha = -1.15 \pm 0.20$.
Regarding these two choices as indicative of the range of systematic
error, we adopt $\alpha = -1.20 \pm 0.25$ as a final result
and uncertainty.

Although there is no strong evidence of additional reddening intrinsic
to the afterglow, one cannot rule out small amounts that are comparable to
the systematic uncertainties in the Galactic value.  An additional constraint
on the total extinction can be obtained in a weakly model-dependent
way by comparing the extrapolated
optical spectrum to the simultaneous X-ray flux as measured by the
BeppoSAX NFI.  If the synchrotron afterglow models have any
validity, then the optical-to-X-ray spectral index $\alpha_{\rm ox}$
should be less than or equal to the $\alpha = -1.20 \pm 0.25$ measured
in the optical band alone.  That is, if there are any breaks in the
broad-band spectrum, they should be concave downward.  In order to make this
comparison, we must estimate the X-ray flux at the fiducial time
of the synthesized optical spectrum.  The X-ray afterglow spectrum is fitted
by energy index $\alpha_{\rm x} = -1.52^{+0.70}_{-0.57}$ (Owens et al. 1998).
Using this index, we convert the flux of 
$(3.8 \pm 0.6) \times 10^{-13}$ ergs~cm$^{-2}$~s$^{-1}$
measured in the 2--10 keV band in the
interval between 9.7 and 13.7 hours after the burst (Nicastro et al. 1998)
to a flux of $0.021\ \mu$Jy at 4.5~keV.  
The X-ray temporal decay of GRB~980518,
$\delta_{\rm x} = -2.07 \pm 0.11$, is the fastest
of the seven afterglows that were well measured by BeppoSAX (Owens et al.
1998).  From this temporal decay, we infer that by $t= 19.8$~h
the flux had faded by a factor of 3 to $0.007\ \mu$Jy.  Figure~4 shows
the broad-band optical to X-ray spectrum at $t = 19.8$~h estimated
in this manner.
We allow for an uncertainty of $\pm 50\%$ on the X-ray flux
in this crude analysis.
The optical spectrum and its uncertainties in extrapolation
are consistent with the
X-ray flux, and jointly they prefer $\alpha_{\rm ox}$ of about --1.05.
We conclude that any intrinsic optical extinction is small,
and that a single power law is
a marginally consistent description of all the available optical through 
X-ray data.  We therefore adopt $\alpha_{\rm ox} = -1.05 \pm 0.10$ as an
observed constraint on models.

\section{Interpretation and Conclusions}

Other bursts with well measured light curves have $\delta$
in the range --1.1 to --1.2 (Bloom et al. 1998c;
Diercks et al. 1998; Reichart et al. 1999).
The GRB~980519 data described here provide the one case of a steep
($t^{-2.05}$) afterglow decay that is well documented at several
frequencies.
Therefore, the continuum spectral shape, together with the temporal
decay, can be used to test afterglow models that relate these two
quantities.  In the simplest form of the external
relativistic blast wave model (M\'eszar\'os \& Rees 1997; Wijers, Rees,
and M\'eszar\'os 1997), electrons accelerated to a power-law energy 
distribution proportional to $E^{-p}$,
are responsible for a decaying synchrotron spectrum of the form
$F_{\nu} \propto \nu^{\alpha}\,t^{\delta}$.
The energy index $\alpha = (1-p)/2$ if the cooling time is longer
than the age of the shock (the adiabatic case), and it is related to
the corresponding temporal decay constant as $\delta = 3\alpha/2$.
Since we observe $\delta = -2.05 \pm 0.04$ all across the
spectrum from optical to X-ray, we would expect a spectral slope
$\alpha_{\rm ox} = -1.37$, incompatible with the observed
$\alpha_{\rm ox} = -1.05 \pm 0.10$.  Although the optical slope alone
is consistent with $\alpha = -1.37$, such a model falls short
of matching the X-ray flux by a factor of 10.
Equivalently, $\alpha = -1.05 \pm 0.10$ would predict
$\delta = -1.58 \pm 0.15$,
incompatible with either the X-ray or the optical decay.
If, instead, we assume that the electrons are in the ``cooling''
regime, for which $\alpha = -p/2$ and $\delta = (3\alpha + 1)/2$
(e.g., Sari, Piran, \& Narayan 1998), then the discrepancy is even
worse, since $\alpha = -1.05$ would require $\delta = -1.08$.
If there is actually a cooling break {\it between} the
optical and the X-ray at the epoch illustrated in Figure~4,
or additional extinction in the GRB host galaxy,
it would exacerbate the discrepancy between $\delta$ and $3\alpha/2$
in the optical.

Of course, these are idealized models involving isotropic
expansion into a homogeneous medium.  One modification to the model
that might be more compatible with GRB~980519 is an inhomogeneous
medium.  M\'eszar\'os, Rees, \& Wijers (1998) calculated the
effects of differing power-law density distributions $n(r)$ for
the shocked medium and concluded that $n(r) \propto r^{-2}$ would
result in $\alpha = -1$ if $\delta = -2$.  Both of these values
are consistent with GRB~980519.
An $n(r) \propto r^{-2}$ dependence is appropriate
for a pre-existing stellar wind.

Alternatively, one can consider anisotropic beaming models.
It has long been hypothesized that GRBs are beamed, as
a way of ameliorating the energetics problem. 
Although we have no redshift for GRB~980519 and therefore
no handle on its energetics, the fact that its host galaxy
at $R = 26.1$ ($R = 25.4$ corrected for Galactic extinction)
is one of the fainter of the nine probable hosts detected so far
means that it could be quite distant and energetic.
[GRB~981220 has a coincident object of $R = 26.4 \pm 0.7$ (Bloom et al. 1999).
Only GRB~980326 is apparently lacking a host galaxy to a limiting
magnitude $R = 27.3$ (Bloom \& Kulkarni 1998).]
Assuming that GRB~980519 is at $z > 1$,
for $H_0 = 65$ km~s$^{-1}$~Mpc$^{-1}$
and $\Omega = 0.2$ its luminosity distance is at least
$2.0 \times 10^{28}$~cm and the BATSE measured fluence
of $(2.54 \pm 0.41) \times 10^{-5}$ ergs~cm$^{-2}$ corresponds to an isotropic
energy of $6.3 \times 10^{52}$~ergs.  This number would rise to
$6.1 \times 10^{53}$~ergs at $z = 3$, greater
than that of the highest redshift burst, GRB~971214 at $z = 3.42$,
for which an isotropic energy of $3 \times 10^{53}$~ergs was inferred
(Kulkarni et al. 1998). 

Jet models (Rhoads 1999) predict a transition from radial
expansion to lateral spreading, after which the temporal decay
steepens to $t^{-p}$.  Values of $p$ in the range 2--2.5 are expected.
In an accompanying paper (Sari et al. 1999), predictions
of jet models for GRB~980519 and
other afterglows are explored, under the assumption that
a transition from radial to transverse expansion had already occurred before
the first optical observation. 
We consider that the combination
of the steep afterglow decay and faint and possibly distant
host galaxy of GRB~980519
make it a good candidate for a jet.

\acknowledgments
 
\clearpage
 
\begin{deluxetable}{llcrr}
\footnotesize
\tablecaption{Optical Photometry of GRB~980519. \label{tbl-1}}
\tablewidth{0pt}
\tablehead{
\colhead{Date (1998 UT)} & \colhead{Telescope}   & \colhead{Filter}   & 
\colhead{Magnitude} &  \colhead{Reference}} 
\startdata
May 19.88  & NOT 2.5m     & I & $18.38 \pm 0.1$  & 1 \nl
May 20.00  & NOT 2.5m     & I & $18.95 \pm 0.03$ & 1 \nl
May 20.31  & MDM 1.3m     & I & $20.05 \pm 0.07$ & 2 \nl
May 20.43  & Palomar 5m   & I & $20.79 \pm 0.11$ & 3 \nl
May 20.98  & NOT 2.5m     & I & $21.54 \pm 0.2$  & 1 \nl
May 21.17  & NOT 2.5m     & I & $21.54 \pm 0.1$  & 1 \nl
May 21.35  & Yerkes 41 in & I & $21.8 \pm 0.3$   & 4 \nl
May 21.43  & Palomar 5m   & I & $>21.5 \pm 0.7$  & 5 \nl
May 20.163 & USNO 40 in   & R & $20.39 \pm 0.12$ & 6 \nl
May 20.229 & USNO 40 in   & R & $20.77 \pm 0.15$ & 6 \nl
May 20.287 & USNO 40 in   & R & $20.87 \pm 0.13$ & 6 \nl
May 20.31  & MDM 1.3m     & R & $20.76 \pm 0.07$ & 2 \nl
May 20.4   & MRO 0.76m    & R & $21.00 \pm 0.25$ & 7 \nl
May 20.44  & APO 3.5m     & R & $21.10 \pm 0.03$ & 8 \nl
May 20.445 & USNO 40 in   & R & $21.15 \pm 0.13$ & 6 \nl
May 20.48  & Palomar 5m   & R & $21.50 \pm 0.09$ & 3 \nl
May 21.469 & Palomar 5m   & R & $23.41 \pm 0.20$ & 5 \nl
May 21.6   & Keck II 10m  & R & $23.03 \pm 0.13$ & 5 \nl
July 18.5  & Keck II 10m  & R & $26.1 \pm 0.3$ & 9 \nl
July 24    & BTA 6m       & R & $26.05 \pm 0.22$ & 10 \nl
May 20.34  & WIYN 3.5m    & V & $21.56 \pm 0.08$ & 2 \nl
May 20.466 & Palomar 5m   & V & $21.94 \pm 0.16$ & 5 \nl
May 21.476 & Palomar 5m   & V & $>22.2$ & 5 \nl
May 20.057 & Wise 1m      & B & $21.09 \pm 0.25$ & 11 \nl
May 20.449 & Palomar 5m   & B & $22.67 \pm 0.14$ & 5 \nl
May 21.448 & Palomar 5m   & B & $> 23.0$ & 5 \nl
May 19.863 & OCA 0.9m     & none & $19.06 \pm 0.26$ & 12 \nl
May 20.077 & OCA 0.9m     & none & $19.81 \pm 0.37$ & 12 \nl
May 20.964 & OCA 0.9m     & none & $22.06 \pm 0.76$ & 12 \nl
\enddata
\tablenotetext{}{References.--(1) Hjorth et al. (1998); (2) this paper;
(3) Bloom et al. (1998b); (4) Castander et al. (1998); (5) Gal et al. (1998);
(6) Vrba et al. (1998); (7) Diercks \& Morgan (1998);
(8) Diercks \& Stubbs (1998); (9) Bloom et al. (1998b);
(10) Sokolov et al. (1998); (11) Leibowitz \& Ibbetson (1998);
(12) Maury et al. (1998).} 
\end{deluxetable}

\clearpage

%
%

\clearpage

\begin{figure}
\plotone{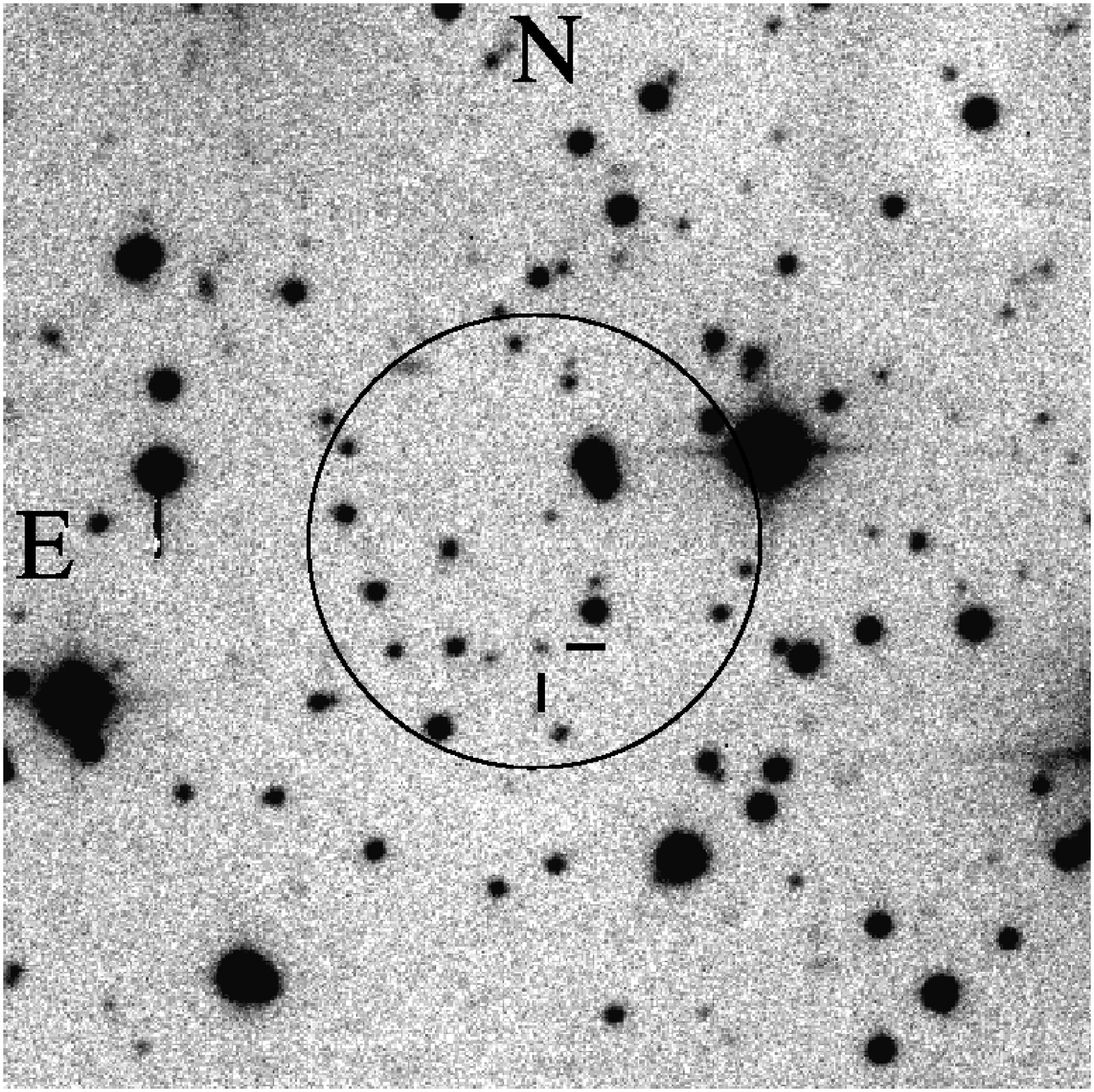}
\caption{``Finding chart'' for the
GRB~980519 optical transient made from the
MDM 1.3m $R$-band image of May 20.31, where $R_{\rm OT} = 20.76$.
A $4^{\prime}\times4^{\prime}$ section of the image is displayed.
The BeppoSAX NFI error circle 
of radius $50^{\prime\prime}$ (Nicastro et al. 1998) is indicated.
The position of the OT is
(J2000) $23^{\rm h}22^{\rm m}21.\!^{\rm s}49,
+77^{\circ}15^{\prime}43.\!^{\prime\prime}3$. \label{fig1}}
\end{figure}

\begin{figure}
\plotone{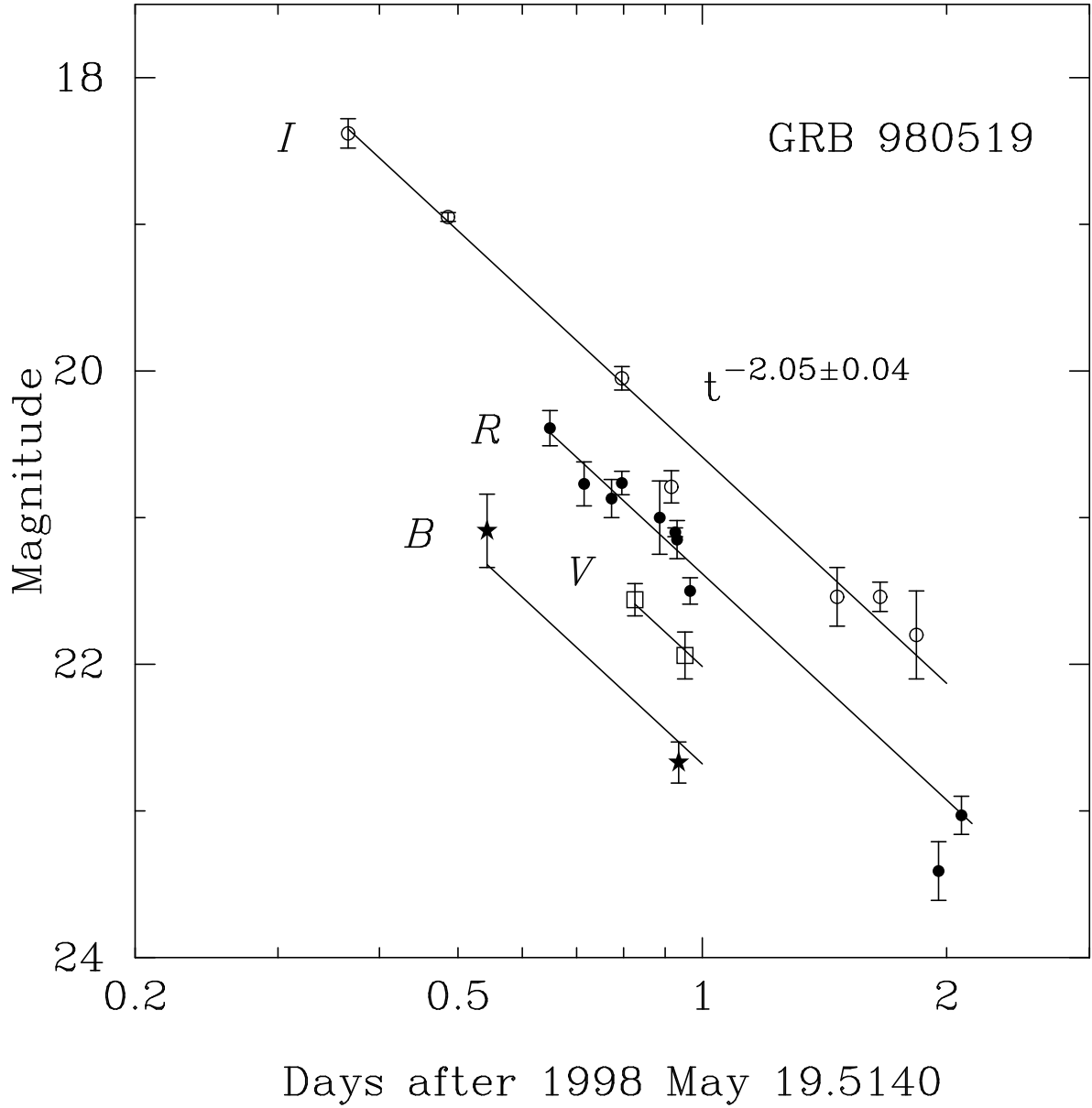}
\caption{Light curves of GRB~980519 in $BRVI$.
The data are taken from Table~1.  Upper limits and unfiltered
observations have been omitted for clarity. \label{fig2}}
\end{figure}

\begin{figure}
\plotone{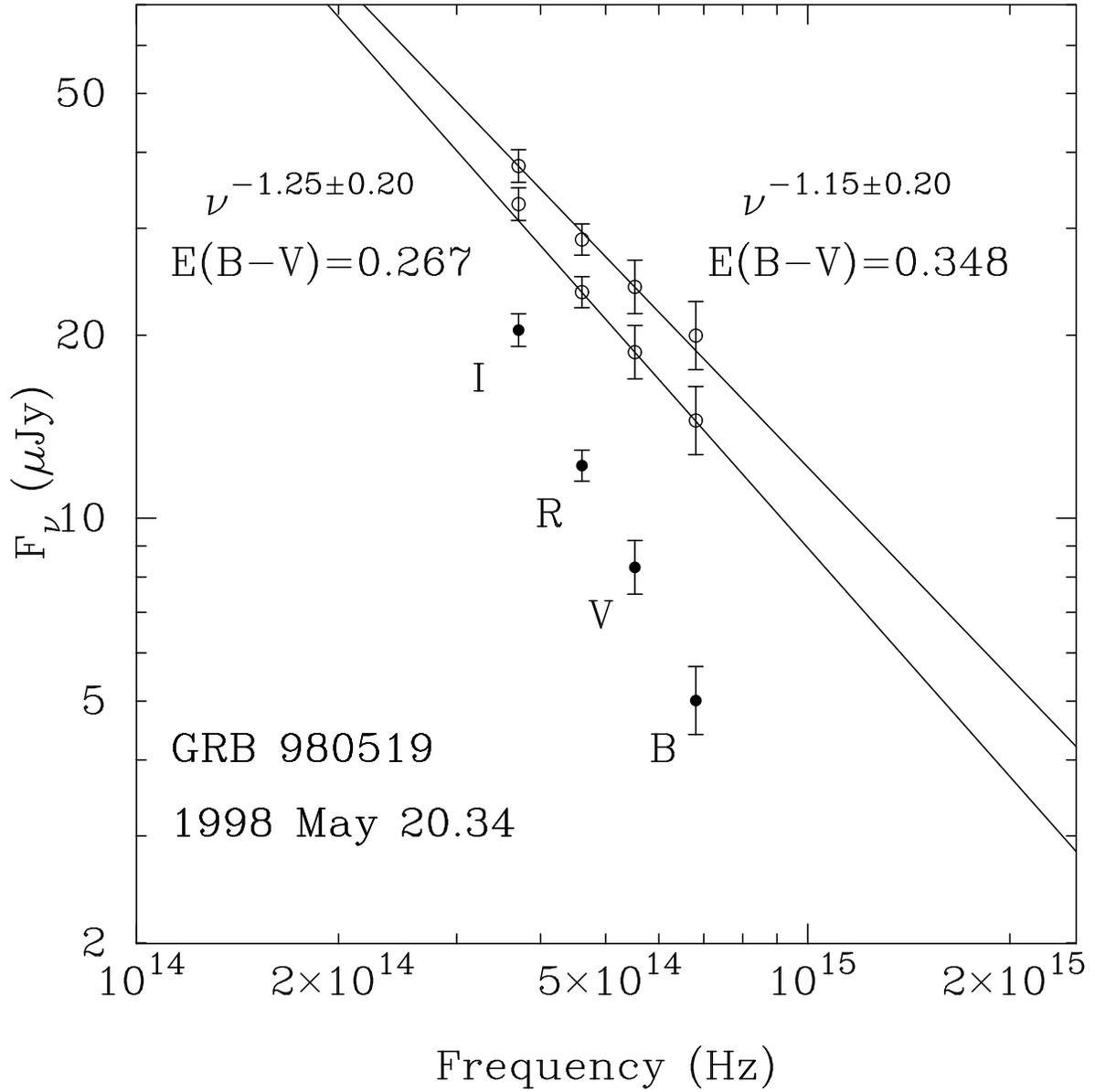}
\caption{Synthetic spectrum of GRB~980519 19.8 h after
the burst, constructed from the fits in Figure~2 (filled circles).
Two different estimates
of the Galactic extinction, as described in the text, are used to
deredden the fluxes (open circles), which leads to slightly
different power-law fits. \label{fig3}}
\end{figure}

\begin{figure}
\plotone{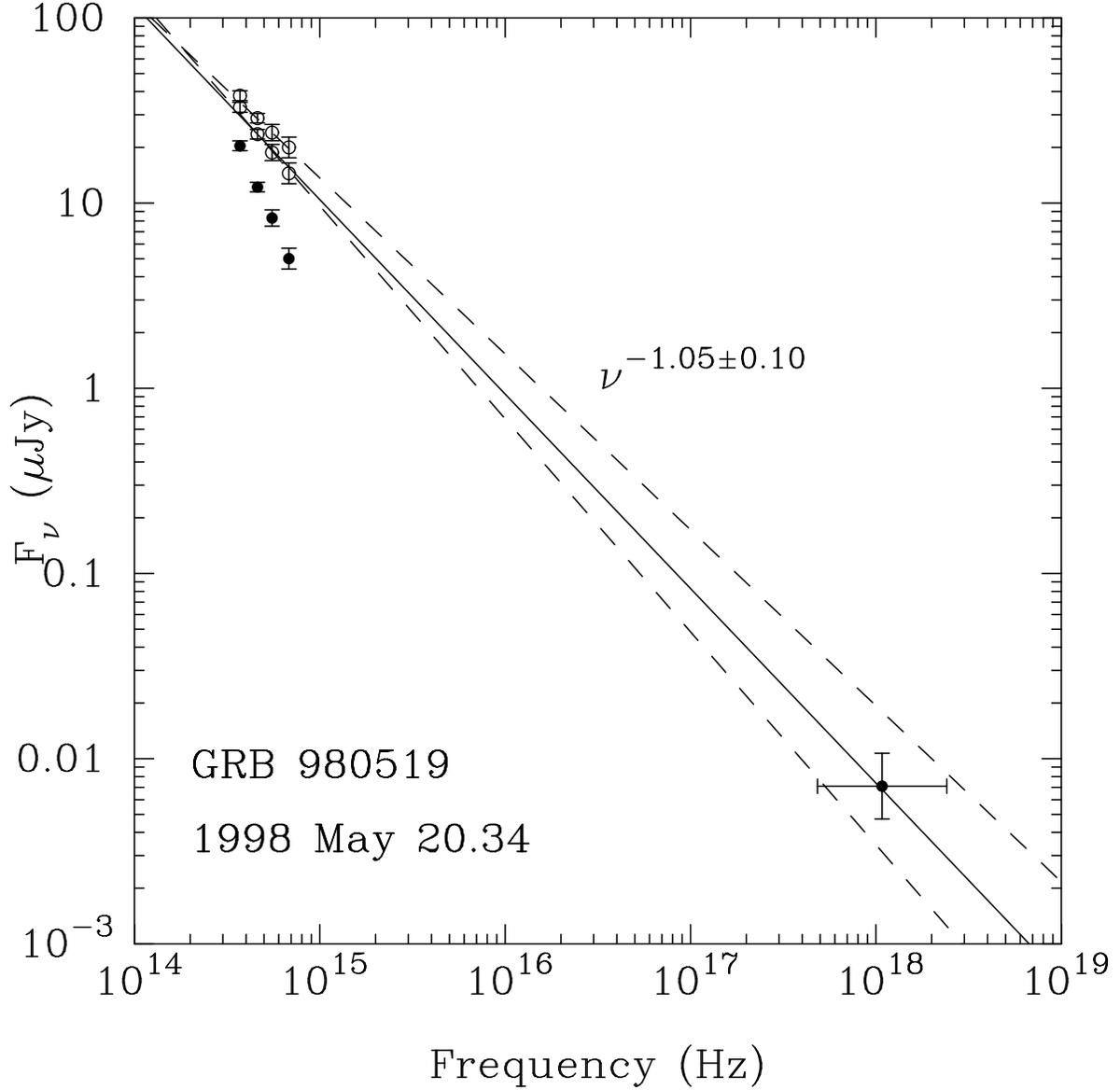}
\caption{A broad-band view of the spectrum of
GRB~980519 19.8 h after the burst, including the BeppoSAX
NFI X-ray flux.  As in Figure 3, two different estimates of
the Galactic extinction are used to deredden the optical fluxes.
The X-ray flux is calculated from the 2--10~keV data in
Nicastro et al. (1998) and Owens et al. (1998), and is assigned
an uncertainty of $\pm50\%$.
The dashed lines represent the estimated uncertainty on the optical-to-X-ray
spectral index, $\alpha_{\rm ox} = -1.05 \pm 0.10$. \label{fig4}}
\end{figure}

\end{document}